\documentclass[prb,showpacs,superscriptaddress,twocolumn]{revtex4-1}
\usepackage{amssymb,amsmath,color}

\begin{document}

\newcommand{\kv}[0]{\mathbf{k}}
\newcommand{\Rv}[0]{\mathbf{R}}
\newcommand{\rv}[0]{\mathbf{r}}
\newcommand{\K}[0]{\mathbf{K}}
\newcommand{\Kp}[0]{\mathbf{K'}}
\newcommand{\dkv}[0]{\delta\kv}
\newcommand{\dkx}[0]{\delta k_{x}}
\newcommand{\dky}[0]{\delta k_{y}}
\newcommand{\dk}[0]{\delta k}
\newcommand{\cv}[0]{\mathbf{c}}
\newcommand{\qv}[0]{\mathbf{q}}

\title{Low frequency optical conductivity in graphene and in other scale-invariant two-band systems}
\author{\'Ad\'am B\'acsi}
\email{bacsi.adam@wigner.bme.hu}
\affiliation{Department of Physics, Budapest University of Technology and Economics, 1521 Budapest, Hungary}
\author{Attila Virosztek}
\affiliation{Department of Physics, Budapest University of Technology and Economics, 1521 Budapest, Hungary}
\affiliation{Institute for Solid State Physics and Optics, Wigner Research Center for Physics, PO Box 49, 1525 Budapest, Hungary}
\date{\today}

\begin{abstract}
We investigate optical transitions of non-interacting electron systems consisting of two symmetric energy bands touching each other at the Fermi energy (e.g. graphene at half filling). Optical conductivity is obtained using Kubo formula at zero temperature. We show that for particles whose pseudospin direction is determined solely by the direction of their momentum, the optical conductivity has power law frequency dependence with the exponent $(d-2)/z$ where $d$ is the dimension of the system and $z$ is the dynamical exponent. According to our result two-dimensional systems with the above pseudospin characteristics always exhibit frequency-independent optical conductivity.
\end{abstract}


\maketitle

\section{Introduction}

Graphene is a two-dimensional material consisting of carbon atoms arranged in a honeycomb lattice. Since its discovery\cite{novoselov1}, it attracted lots of attention due to its peculiar electronic and optical properties\cite{peresrevmod,sarmarevmod}. Graphene is a promising material in industrial applications and is also very important from theoretical point of view because its quasiparticles can be described as massless Dirac fermions. These particles possess pseudospin degree of freedom originating from the two atoms of the unit cell. 
One of the most fascinating properties of monolayer graphene is that the optical conductivity is universal and independent of the frequency\cite{unicondfalkovsky,unicondmarel,unicondgeim,optresmeas}. This feature was argued to be a consequence of the linear spectrum and the structure of the Dirac cones\cite{sharapovoptcond}.

Electronic and optical properties of multilayer graphene have also been investigated experimentally and theoretically \cite{thz2,optmeas2,peresmulti}. The band structure of multilayer graphene is very sensitive to its stacking sequence \cite{MM08}. In the case of periodic ABC stacking, the low energy behaviour of quasiparticles show chiral nature and can be characterized by the pseudospin winding number which is equal to the number of layers. Chiral nature of quasiparticles has been proven to play an important role in various phenomena, such as quantum Hall effect \cite{qhe}, THz radiation measurements \cite{thz} and angle-resolved photoemission spectroscopy measurements \cite{arpes}.

In periodic ABC stacked (sometimes called chiral or orthorhombic) multilayer graphene the Hamiltonian can be written as
\begin{equation}
\label{eq:ham}
\hat{H}_{\nu}(\kv)\sim\nu k^{M}\left[\begin{array}{cc} 0 & e^{-iM\nu\gamma} \\ e^{iM\nu\gamma} & 0 \end{array} \right]
\end{equation}
where $k$ and $\gamma$ are the magnitude and the angle of the vector $\kv$ which is the wavenumber measured from the corners of the Brillouin zone. In the formula, $M$ is the number of layers and $\nu=\pm 1$ is the valley index describing the two inequivalent corners of the Brillouin zone. The optical conductivity of ABC stacked multilayer graphene is also found to be universal and independent of frequency\cite{mdoptresmlg}.

In the present paper the optical conductivity of two-band systems is studied within Kubo formalism. We review earlier calculations\cite{csertizitt} and obtain the real part of the optical conductivity expressed in terms of the pseudospin of the particles. Analytical results are obtained for two-band systems in which the pseudospin of the quasiparticles is determined only by the direction of their momentum (as well as in the case of chiral multilayer graphene). The optical conductivity of general chiral systems defined in this way exhibits power-law frequency dependence. This behaviour is a consequence of the fact that there is no energy scale in these systems, therefore, scaling arguments can be applied.

We show that the universal behaviour of the optical conductivity in chiral multilayer graphene can be explained by the two-dimensionality and the invariance under dilatations.


\section{Optical conductivity in solids}
Physical quantities measured in optical experiments can usually be expressed by means of the optical conductivity. 
Using the Kubo formula the optical conductivity can be written as
\begin{equation}
\label{eq:kubo1}
\sigma_{\alpha\beta}(\rv,\rv',\omega)=-\frac{e^{2}n(\rv)}{i\omega m}\delta(\rv-\rv')\delta_{\alpha\beta}+$$$$+\frac{1}{i\omega}\lim_{\delta\rightarrow 0^{+}}\int_{0}^{\infty} dt\, e^{i\omega t-\delta t}\frac{i}{\hbar}\langle[j_{\alpha}(\rv,t),j_{\beta}(\rv',0)]\rangle_{0}
\end{equation}
where $n(\rv)$ is the density of electrons, $j_{\alpha}(\rv,t)$ is the current density operator and the expectation value shall be evaluated in the equilibrium state. The first term in (\ref{eq:kubo1}) is the diamagnetic term. Henceforth, we neglect the notation $\lim_{\delta\rightarrow 0^{+}}$.

The one-particle eigenstates of a solid are the Bloch states which obey the $H(\rv)\Psi_{l\kv}(\rv)=E_{l}(\kv)\Psi_{l\kv}(\rv)$ Schr\" odinger equation where $H(\rv)=-\hbar^{2}\Delta/(2m)+U(\rv)$ with lattice periodic potential. Because of Bloch's theorem the eigenfunctions can be rewritten as $\Psi_{l\kv}(\rv)=e^{i\kv\rv}u_{l\kv}(\rv)/\sqrt{\Omega}$ where $u_{l\kv}(\rv)$ is a periodic function of space variables and $\Omega$ is the volume of the sample.
If the Coulomb interaction between electrons is neglected then one-particle eigenstates are filled following the Fermi distribution function $f(E)$.

The $\qv=0$ Fourier component of the frequency dependent conductivity
$$\sigma_{\alpha\beta}(\omega)=\sigma^{dia}_{\alpha\beta}(\omega)-\frac{1}{i\omega}\frac{2e^{2}}{\Omega}\sum_{ll'\kv\kv'}V_{\alpha,ll'\kv\kv'}V_{\beta,l'l\kv'\kv}\times$$
\begin{equation}
\label{eq:kubo2}
\times
\frac{f(E_{l}(\kv))-f(E_{l'}(\kv'))}{\hbar\omega+i\delta+E_{l}(\kv)-E_{l'}(\kv')}
\end{equation}
where $\kv$ and $\kv'$ run over the whole Brillouin zone. The matrix elements $\mathbf{V}_{ll'\kv\kv'}=\langle\Psi_{l\kv}|\mathbf{v}|\Psi_{l'\kv'}\rangle$ of the velocity operator may be rewritten as
\begin{equation}
\label{eq:velocity}
\mathbf{V}_{ll'\kv\kv'}=\delta_{\kv\kv'}\left[i\omega_{ll'}(\kv)\mathbf{A}_{ll'}(\kv)+\mathbf{v}_{l}(\kv)\delta_{ll'}\right]
\end{equation}
with $\hbar\omega_{ll'}(\kv)=E_{l}(\kv)-E_{l'}(\kv)$. In the formula, $\hbar\mathbf{v}_{l}(\kv)=\partial E_{l}/\partial\kv$ is the quasiparticle velocity in the $l$th band and
\begin{equation}
\label{eq:conn}
\mathbf{A}_{ll'}(\kv)=\frac{i}{\Omega}\int d^{d}r\,u^{*}_{l\kv}(\rv)\nabla_{\kv}u_{l'\kv}(\rv)
\end{equation}
is the self-adjoint Berry connection matrix.
Note that the first term in (\ref{eq:velocity}) describes transitions between two different bands and equals to zero if $l=l'$. The second term is non-zero only if $l=l'$ and is the quasiparticle velocity as expected. However, in (\ref{eq:kubo2}) only terms corresponding to two different bands survive for non-zero frequencies because of the Fermi functions. Therefore,

$$\sigma_{\alpha\beta}(\omega)=\sigma_{\alpha\beta}^{dia}(\omega)-\frac{2e^{2}}{i\hbar\omega}\frac{1}{\Omega}\sum_{ll'\kv}\omega_{ll'}(\kv)^{2}\times$$
\begin{equation}
\label{eq:kubo3}
\times A_{\alpha,ll'}(\kv)A_{\beta,l'l}(\kv)\frac{f(E_{l}(\kv))-f(E_{l'}(\kv))}{\omega+i\delta+\omega_{ll'}(\kv)}\,.
\end{equation}

In the followings, we focus on the real part of the optical conductivity. The imaginary part may be calculated using Kramers-Kronig relations. The diamagnetic conductivity is purely imaginary so it does not appear in the real part at non-zero frequencies. The $\delta\rightarrow 0^{+}$ limit of the second term in (\ref{eq:kubo3}) can be divided into a principal value part and a Dirac-delta part. Since the Hamiltonian does not depend on the spin, the system is time reversal invariant. This leads to $E_{l}(\kv)=E_{l}(-\kv)$ and $\mathbf{A}_{ll'}(\kv)=\mathbf{A}_{l'l}(-\kv)$ and, hence, the real part of the conductivity equals to the Dirac-delta part.
We note that this statement is true even if the Hamiltonian is spin-dependent, for example in the presence of magnetic field. In this case the statement is a consequence of $E_{l}^{s}(\kv)=E_{l}^{-s}(-\kv)$ and $\mathbf{A}^{s}_{ll'}(\kv)=\mathbf{A}^{-s}_{l'l}(-\kv)$ with $s$ denoting the different spin directions, but magnetic effects are out of scope of the present paper.
Therefore,
$$
\mbox{Re}\,\sigma_{\alpha\beta}(\omega)=\frac{2e^{2}}{\hbar}\frac{\pi\omega}{\Omega}\sum_{ll'\kv}A_{\alpha,ll'}(\kv)A_{\beta,l'l}(\kv)\times
$$
\begin{equation}
\label{eq:opt1}
\times
\big[f(E_{l}(\kv))-f(E_{l'}(\kv))\big]\delta(\omega-\omega_{l'l}(\kv))
\end{equation}
can be obtained for non-zero frequencies. Note that intraband processes do not contribute for non-zero frequencies\footnote{In the presence of disorder it is true for frequencies larger than the scattering rate.}. The optical conductivity is now expressed in terms of the Berry connection matrix.
It is well known that the Berry connection is essential in determining the $dc$ Hall conductivity\cite{thoulesshall}. However, Eq. (\ref{eq:opt1}) shows that this topological quantity plays an important role in the optical conductivity as well.

Using Eq. (\ref{eq:opt1}) one can calculate the optical conductivity of solids generally. To do so, first the energy spectrum and the Berry connection matrix have to be computed in band structure calculations. In tight binding approximation the Hilbert space is restricted to several atomic orbits per unit cell and the eigenfunctions are given as $e^{i\kv\rv}|\kv,l\rangle/\sqrt{\Omega}$ where $|\kv,l\rangle$ is a finite-dimensional vector representing the eigenstate on this subspace. In order to calculate the Berry connection matrix elements one may use $\mathbf{A}_{ll'}(\kv)=i\langle\kv,l|\nabla_{\kv}|\kv,l'\rangle$ instead of (\ref{eq:conn}). This expression depends on the choice of the basis in the tight binding subspace but we claim that in most cases $i\langle\kv,l|\nabla_{\kv}|\kv,l'\rangle$ deviates from (\ref{eq:conn}) negligibly in the "natural" tight binding basis. For details, see Appendix.

\section{Two band system}
For frequencies lower than the bandwidth, only optical transitions between the two bands closest to the Fermi level are allowed. 
We focus on systems with two bands where the lower band is fully occupied and the upper band is empty and we consider zero temperature.

The Hamiltonian of two band models can be represented on this subspace with a self-adjoint $2\times 2$ matrix.
Assuming that it is independent of the spin the Hamilton operator can generally be written as $$\hat{H}(\kv)=\varepsilon_{0}(\kv)\hat{I}+\varepsilon(\kv)\vec{n}(\kv)\hat{\vec{\sigma}}$$
where $\varepsilon_{0}(\kv)$ and $\varepsilon(\kv)\geq 0$ are real functions.
The three-dimensional vector $\hat{\vec{\sigma}}$ is the pseudospin operator built up from the Pauli matrices and $\hat{I}$ is the $2\times 2$ identity matrix. The components of the three-dimensional unit vector $\vec{n}(\kv)$ are real functions of the wavenumber. 

Due to time reversal invariance of the Hamiltonian $\varepsilon(-\kv)=\varepsilon(\kv)$, $\varepsilon_{0}(-\kv)=\varepsilon_{0}(\kv)$ and $\{n_{x}(-\kv),n_{y}(-\kv),n_{z}(-\kv)\}=\{n_{x}(\kv),-n_{y}(\kv),n_{z}(\kv)\}$. 
The energy spectrum of the two bands is $E_{\pm}(\kv)=\varepsilon_{0}(\kv)\pm \varepsilon(\kv)$ where the $+$ and $-$ signs correspond to the conductance and valance band, respectively.
Since $\vec{n}$ is a three-dimensional unit vector, it may be described by its azimuthal and polar angles which are also real functions of the wavenumber.
$$n_{x}(\kv)=\sin\vartheta(\kv)\cos\varphi(\kv)$$
$$n_{y}(\kv)=\sin\vartheta(\kv)\sin\varphi(\kv)$$
$$n_{z}(\kv)=\cos\vartheta(\kv)$$
Because of the time reversal properties of the Hamitonian $\vartheta(\kv)$ is an even function while $\varphi(\kv)$ is an odd function of the wavenumber.
The eigenvectors can be written as
$$|\kv,+\rangle=\left[\begin{array}{c} \cos\frac{\vartheta(\kv)}{2} \\ \sin\frac{\vartheta(\kv)}{2}e^{i\varphi(\kv)} \end{array}\right]$$
$$
|\kv,-\rangle=
\left[\begin{array}{c} \sin\frac{\vartheta(\kv)}{2}e^{-i\varphi(\kv)} \\ -\cos\frac{\vartheta(\kv)}{2} \end{array}\right]$$
and are evidently determined by $\vec{n}(\kv)$. It is worth mentioning that $\langle\kv,\pm|\hat{\vec{\sigma}}|\kv,\pm\rangle=\pm\vec{n}(\kv)$, i.e., the vector $\vec{n}(\kv)$ determines the pseudospin of the electrons and holes.

At half filling photons excite electrons from the valance band to the conduction band. These transitions determine the optical conductivity which can be obtained as
\begin{equation}
\label{eq:opt2}
\mbox{Re}\,\sigma_{\alpha\beta}(\omega)=\frac{2e^{2}}{\hbar}\frac{\pi\omega}{\Omega}\sum_{\kv}A_{\alpha,-+}(\kv)A_{\beta,+-}(\kv)
\delta\left(\omega-\frac{2\varepsilon(\kv)}{\hbar}\right)
\end{equation}
for positive frequencies. For negative frequencies $\mbox{Re}\,\sigma_{\alpha\beta}(\omega<0)=\mbox{Re}\,\sigma_{\alpha\beta}(|\omega|)$. Eq. (\ref{eq:opt2}) is valid only at zero temperature. For finite temperature one can give an analytical result only if $\varepsilon_{0}(\kv)= 0$. In this case (\ref{eq:opt2}) has to be multiplied by $\mbox{th}(\hbar\omega/4k_{B}T)$.
Note that $\varepsilon_{0}(\kv)$ does not appear in zero temperature conductivity. 

In tight binding approximation the Berry connection matrix element can be obtained as
\begin{equation}
\label{eq:conn2}
\mathbf{A}_{-+}(\kv)=e^{i\varphi(\kv)}\left[\frac{\nabla_{\kv}\vartheta(\kv)}{2i}+\sin\vartheta(\kv)\frac{\nabla_{\kv}\varphi(\kv)}{2}\right] \,.
\end{equation}

Substituting (\ref{eq:conn2}) into (\ref{eq:opt2}) and taking advantage of the time reversal properties of $\varepsilon(\kv)$, $\vartheta(\kv)$ and $\varphi(\kv)$ we found the following short expression for the optical conductivity. For comparison, see \cite{csertizitt}.
\begin{equation}
\label{eq:opt3}
\mbox{Re}\,\sigma_{\alpha\beta}(\omega)=\frac{e^{2}}{\hbar}\frac{\pi\omega}{2\Omega}\sum_{\kv}\frac{\partial \vec{n}}{\partial k_{\alpha}}\frac{\partial \vec{n}}{\partial k_{\beta}}\delta\left(\omega-\frac{2\varepsilon(\kv)}{\hbar}\right)
\end{equation}

This result shows that the optical conductivity of two-band systems can be calculated by identifying the vector $\vec{n}(\kv)$ and the dispersion $\varepsilon(\kv)$ in the Hamiltonian and then evaluating the sum over the Brillouin zone given in (\ref{eq:opt3}). 
The off-diagonal elements of the optical canductivity do not necessarily vanish and describe Faraday rotation if the symmetry group of the system is low enough.

Applying (\ref{eq:opt3}) to chirally stacked multilayer graphene (\ref{eq:ham}) and taking into account both valleys, we obtain the well-known universal result
\begin{equation}
\label{eq:gr}
\sigma_{\alpha\beta}^{\mathrm{multiLG}}(\omega)=\frac{M}{4}\frac{e^{2}}{\hbar}\delta_{\alpha\beta}
\end{equation}
which is independent of the frequency\cite{mdoptresmlg}. We note that the low energy model of multilayer graphene (\ref{eq:ham}) is time reversal invariant only if both valleys are taken into account.

In the case of monolayer graphene the result (\ref{eq:gr}) was argued to be a consequence of the structure of the Dirac cones and the linear spectrum\cite{sharapovoptcond}. In the next section we show that the universal frequency dependence of the optical conductivity can be found in a wide class of two-band systems.

\section{Scale invariant systems}
The optical conductivity (\ref{eq:opt3}) cannot be calculated generally. However, in this section we define a wide class of two-band systems for which the frequency dependence of the optical conductivity can be evaluated analytically.

We use spherical coordinates in momentum space, i.e., $k$ as the magnitude and $\{\gamma\}$ as the set of angle variables of the wavenumbers. If the dimension of the system is $d$ the set $\{\gamma\}$ consists of $d-1$ angle variables $\{\gamma_{1},\gamma_{2},\dots,\gamma_{d-1}\}$. The gradient in momentum space can be written as 
$$\nabla_{\kv}=\mathbf{e}_{k}\frac{\partial}{\partial k}+\frac{1}{k}\sum_{j=1}^{d-1}c_{j}(\{\gamma\})\mathbf{e}_{j}\frac{\partial}{\partial \gamma_{j}}$$
where $\mathbf{e}_{k}$ and $\mathbf{e}_{j}$ are unit vectors corresponding to the spherical coordinates $k$ and $\gamma_{j}$, respectively. Here $c_{j}$ are coefficients of the Jacobian. Note that if the gradient acts on a function which depends on angle variables only the derivative is propotional to $k^{-1}$.

In this section we focus on two band systems which can be described by the Hamilton operator with the following properties.
We do not consider $\varepsilon_{0}(\kv)$ because, as we have seen in the previous section, it does not influence the zero temperature optical conductivity. Additionally, we assume that the Hamiltonian depends on the magnitude of the momentum only through the dispersion and this dependence is power-law, i. e., $\varepsilon(\kv)=C(\{\gamma\})k^{z}$ where $z$ is the dynamic exponent and $C(\{\gamma\})>0$. In this case $\vec{n}(\kv)=\vec{n}(\{\gamma\})$ is a function of angle variables only.

This assumption has lots of consequences. First, it follows that the conductance and valance bands touch each other at the Fermi level, i.e., there is no energy gap between them. Second, the system is invariant under dilatations which act on spacetime as $\rv\rightarrow b\rv$ and $t\rightarrow b^{z}t$ where $b>0$ is the scaling parameter. This also means that there is no characteristic energy or length scale in these systems.
Third, the eigenvectors depend on angle variables only so the direction of the pseudospin is determined by the direction of the momentum only. 
Since this feature is a kind of generalization of the chirality of neutrinos and electrons in graphene, from now on, we may also refer to systems with the above characteristics as general chiral systems.
Note that monolayer graphene and chiral stacked multilayer graphene are general chiral systems.  

The density of states in general chiral systems is a power-law function of the energy.
$$g(\varepsilon)=\frac{\kappa}{z\pi}|\varepsilon|^{\frac{d}{z}-1}$$
\begin{equation}
\label{eq:dos}
\kappa=\int\frac{d\,\{\gamma\}}{(2\pi)^{d-1}}C(\{\gamma\})^{-\frac{d}{z}}
\end{equation}
where the notation $d\,\{\gamma\}$ stands for the integration with respect to all angle variables and the Jacobian determinant is also incorporated. In mean-field theory systems with power-law density of states exhibit interesting critical behaviour in quantum phase transitions accompanied by gap-opening\cite{quantummf}.

In order to calculate the optical conductivity one has to determine the derivatives $\partial\vec{n}/\partial k_{\alpha}$. Since $\vec{n}(\{\gamma\})$ is a function of angle variables only this derivative is proportional to $k^{-1}$.
In the thermodynamic limit the summation in Eq. (\ref{eq:opt3}) becomes an integral over the momentum space. This integral can be factorized into an integral with respect to angle variables and an integral with respect to $k$. Similarly to calculations presented in Ref. \cite{mdoptresmlg}, the latter integral can be carried out analytically in the case of general chiral systems and also determines the frequency dependence of the optical conductivity.
\begin{equation}
\label{eq:scf}
\mbox{Re}\,\sigma_{\alpha\beta}(\omega)=\frac{e^{2}}{\hbar}\frac{K_{\alpha\beta}}{4z}\kappa^{\frac{d-2}{d}}\left(\frac{\hbar\omega}{2}\right)^{\frac{d-2}{z}}
\end{equation}
where $\kappa$ is the coefficient of the density of states (\ref{eq:dos}) and $K_{\alpha\beta}$ is a dimensionless coefficient defined as
$$K_{\alpha\beta}=\kappa^{\frac{2-d}{d}}\int \frac{d\,\{\gamma\}}{(2\pi)^{d-1}} C(\{\gamma\})^{\frac{2-d}{z}}\times$$$$\times
\sum_{jj'=1}^{d-1}c_{j}(\{\gamma\})e_{j,\alpha}\frac{\partial\vec{n}}{\partial \gamma_{j}}c_{j'}(\{\gamma\})e_{j',\beta}\frac{\partial\vec{n}}{\partial \gamma_{j'}}\,.$$

We note that in certain cases (for example in graphene) Eq. (\ref{eq:scf}) has to be multiplied by valley degeneracy.
Since neither $\kappa$ nor $K_{\alpha\beta}$ depend on the frequency, the optical conductivity has universal $\omega^{(d-2)/z}$ frequency dependence. 
Using Kramers-Kronig relations the imaginary part of the conductivity can be obtained as $\mbox{Im}\,\sigma(\omega)\sim \omega^{(d-2)/z}$ at low frequencies.
In a two-dimensional system (\ref{eq:scf}) provides a frequency-independent result. 

In one dimension there are no angle variables so the unit vector $\vec{n}$ does not depend on the wavenumber which means that its derivative vanishes. It follows that in one-dimensional general chiral systems the optical conductivity is identically zero for finite frequency.

The result (\ref{eq:scf}) is universal in the sense that the microscopic details do not influence the frequency dependence but are incorporated in prefactors only. The universal behaviour can be understood on the basis of dilatation invariance. The optical conductivity scales as $\sigma(b,\omega)=b^{2-d}\sigma(1,b^{z}\omega)$ under dilatations defined above and the dilatation invariance demands $d\sigma/db=0$. This equation results in $\sigma(\omega)\sim\omega^{(d-2)/z}$ but says nothing about the prefactors.

Dynamic scaling analysis is a very powerful tool to explore the frequency dependence of the optical conductivity in the vicinity of the critical point of phase transitions\cite{dynscale1,unicondqpt,dynscale2}. Exactly at $T_{c}$, power-law dependence with the exponent $(d-z-2)/z$ has been found \cite{dynscale1}. In quantum phase transitions one may use the effective dimension $d_{eff}=d+z$ instead of $d$ which leads to the same frequency dependence as in Eq. (\ref{eq:scf}). It follows that in two-dimensional quantum critical systems the optical conductivity is also frequency-independent \cite{unicondqpt}.

The scaling argument can be applied to other physical quantities. For example, the static polarization function of general chiral systems depends on the wavenumber as $\chi(q)\sim q^{d-z}$ in agreement with \cite{dassarmapol}.
Note that the scale invariance is closely related to the fact that there is no characteristic energy in the system. If an energy scale appears, for example due to gap-opening, Eq. (\ref{eq:scf}) is not valid any more and the frequency dependence will be influenced by microscopic details, for example the way how the gap has been formed.

\section{Conclusion}
We presented a calculation of low frequency optical conductivity based on Kubo formalism. As a result, we obtained universal frequency dependence in systems which are invariant under dilatations and, hence, do not have characteristic energy scale. Scale invariance is closely related to the fact that the pseudospin is determined by the direction of the momentum only. One may think of this property as some kind of chirality in a very general sense. The exponent of the frequency dependence is determined solely by the dynamic exponent and the dimension of the system. In two dimensions the optical conductivity is  independent of the frequency for any dynamic exponent in any scale invariant system.

\begin{acknowledgments}
We acknowledge fruitful discussions with Bal\'azs D\'ora. This work was supported by the Hungarian Scientific Research Fund under Grants No. OTKA K101244.
\end{acknowledgments}

\bibliographystyle{apsrev}
\bibliography{optchiral}

\appendix
\section{Berry connection in tight-binding approximation}
The elements of Berry connection matrix are defined in (\ref{eq:conn}). However, there is no analytical method to calculate $u_{l\kv}$, one has to apply some kind of approximation.

In this appendix we investigate how the Berry connection matrix can be expressed within tight-binding (TB) approximation.
The localization properties of Wannier functions has been studied extensively\cite{wannier}.

We restrict the Hilbert space to $L$ atomic orbitals $\varphi_{b}(\rv)$ $(b=1,\dots,L)$ per unit cell.
Then one can define the Bloch functions
\begin{equation}
\label{eq:tbbasis}
\varphi_{b\kv}(\rv)=\frac{1}{\sqrt{N}}\sum_{\Rv}e^{i\kv\Rv}\varphi_{b}(\rv-\Rv)
\end{equation}
where $\Rv$ runs over all lattice vectors.  We call this set of functions "natural" TB basis. Note that TB basis is not necessarily orthonormal but can be L\"owdin-orthonormalized\cite{wannier}. We now assume that
$$\int d\rv\varphi_{b\kv}(\rv)^{*}\varphi_{b'\kv'}(\rv)=\delta_{\kv\kv'}\delta_{bb'}\,.$$
The Hamiltonian $H(\rv)=-\hbar^{2}\Delta/(2m)+U(\rv)$ acts on (\ref{eq:tbbasis}) as
$$H(\rv)\varphi_{b\kv}(\rv)=\sum_{b'=1}^{L}H_{b'b}(\kv)\varphi_{b'\kv}(\rv)+(H-PHP)\varphi_{b\kv}(\rv)$$
where $H_{b'b}(\kv)=\int d\rv \varphi_{b'\kv}(\rv)^{*}H(r)\varphi_{b\kv}(\rv)$ and $P=\sum_{b}|\varphi_{b\kv}\rangle\langle\varphi_{b\kv}|$ is a projector onto the subspace spanned by the TB basis.
In tight-binding approximation $(H-PHP)\varphi_{b\kv}(\rv)$ is neglected.

In practice, the starting point in TB calculations is to determine $H_{b'b}(\kv)$.
Then, the solutions of the Schr\"odinger equation can be given as $\varphi_{l\kv}(\rv)=\sum_{b}c_{lb}(\kv)\varphi_{b\kv}(\rv)$ where the coefficients $c_{lb}(\kv)$ are determined by the eigenvectors of $H_{b'b}(\kv)$.
$$\sum_{b'}H_{b'b}(\kv)c_{lb'}(\kv)=E_{l\kv}c_{lb}(\kv)$$
The periodic part of $\varphi_{l\kv}(\rv)$ can be written as 
$$u_{l\kv}(\rv)=\sqrt{\Omega}e^{-i\kv\rv}\varphi_{l\kv}(\rv)\,.$$
Substituting in (\ref{eq:conn}) we obtain
$$\mathbf{A}_{ll'}(\kv)=i\sum_{b}c_{lb}(\kv)\nabla_{\kv}c_{l'b}(\kv)+$$
\begin{equation}
\label{eq:conntb}
+\sum_{bb'}c_{lb}(\kv)^{*}c_{l'b'}(\kv)\sum_{\Rv}e^{-i\kv\Rv}\int d\rv\,\varphi_{b}(\rv-\Rv)^{*}\rv\varphi_{b'}(\rv)\,.
\end{equation}
By introducing the notation $|\kv,l\rangle$ for the $L$-dimensional vector built up from the coefficients $c_{lb}(\kv)$ the first term in (\ref{eq:conntb}) can be rewritten as $i\langle\kv,l|\nabla_{\kv}|\kv,l\rangle$. In the second term the matrix elements of the dipole operator appear which are taken between Wannier functions. Here we argue that these matrix elements are usually negligible due to symmetry of atomic orbits or their well-localized behaviour but there are a few cases when it is not true. For instance, if both $s$ and $p$ orbits of an atom are taken into account in the tight-binding approximation the matrix element of the dipole operator is not negligible between them. Nevertheless, in most cases (for example in graphene) the second term in (\ref{eq:conntb}) can be neglected and, hence,
$$\mathbf{A}_{ll'}(\kv)=\frac{i}{\Omega}\int d\rv u_{l\kv}(\rv)^{*}\nabla_{\kv}u_{l'\kv}(\rv)
\approx i\langle\kv,l|\nabla_{\kv}|\kv,l\rangle\,.$$


\end{document}